\documentclass[11pt]{article}
\pdfoutput=1
\usepackage{amssymb}
\usepackage{amsmath}
\usepackage{amstext}
\usepackage{graphicx,epsfig}
\usepackage{verbatim} 

\usepackage[margin=1.2in]{geometry}

\newcommand{\be}{\begin{equation}}
\newcommand{\ee}{\end{equation}}
\newcommand{\bea}{\begin{eqnarray}}
\newcommand{\eea}{\end{eqnarray}}
\newcommand{\beas}{\begin{eqnarray*}}
\newcommand{\eeas}{\end{eqnarray*}}

\newcommand{\half}{\frac{1}{2}}

\begin{document}

\long\def\symbolfootnote[#1]#2{\begingroup%
\def\thefootnote{\fnsymbol{footnote}}\footnote[#1]{#2}\endgroup}
{\phantom .}
\begin{center}
\textbf {\Large Smooth Initial Conditions from Weak Gravity} \\
\vspace{1.5em}
Brian Greene,$^{1,2}$\symbolfootnote[1]{{\bf email:} {\it greene@phys.columbia.edu}} Kurt Hinterbichler,$^{1,3}$\symbolfootnote[2]{{\bf email:} {\it kurth@phys.columbia.edu}} Simon Judes,$^1$\symbolfootnote[3]{{\bf email:} {\it judes@phys.columbia.edu}} and Maulik Parikh$^4$\symbolfootnote[4]{{\bf email:} {\it parikh@iucaa.ernet.in}}\\
\vspace{0.75em}
$^1$Institute for Strings, Cosmology and Astroparticle Physics, Department of Physics, Columbia University, New York, NY 10027. \\
\vspace{0.4em}
$^2$Department of Mathematics, Columbia University, New York, NY 10027. \\
\vspace{0.4em}
$^3$Center for Particle Cosmology, Department of Physics and Astronomy, University of Pennsylvania, Philadelphia, PA 19104. \\
\vspace{0.4em}
$^4$ Inter-University Centre for Astronomy and Astrophysics,
Post Bag 4, Pune 411007, India.
\end{center}
\normalsize

\vspace{1ex}
\begin{abstract}
\noindent 
CMB measurements reveal an unnaturally smooth early universe.  We propose a mechanism to make this smoothness natural by weakening the strength of gravity at early times, and therefore altering which initial conditions have low entropy.
\end{abstract}

\vspace{2ex}

\numberwithin{equation}{section}

\newpage
\section{Introduction}
\subsection{The arrow of time}
\ \ \ \ \ \ 

One of the most self--evident features of our experience is the disparity between time evolution of macroscopic systems toward the future and toward the past.  So varied and pervasive are time--asymmetric phenomena, that it is a remarkable fact and an extraordinary example of scientific unification that the vast majority of them can be described in a single statement: entropy increases with time.  This law --- the second law of thermodynamics --- is a powerful result, but it raises some troubling questions.  Boltzmann's entropy is a function of microscopic degrees of freedom, yet the laws that govern microscopic systems possess symmetries that reverse the direction of time.  Where does the asymmetry in time come from?

The natural place to begin is Boltzmann's H--theorem, which purports to derive the second law from classical kinetic theory \cite{Boltzmann}.  The status of the theorem is less settled than often claimed, because it requires the so--called `molecular chaos' assumption, doubts about whose applicability have not been firmly laid to rest.  But it is uncontroversial that one way or another the second law follows from microscopic physics and an appropriate probability distribution over initial microstates.  On the other hand, the existence of a low entropy state to begin with is more puzzling.  Low entropy initial conditions are unnatural in the sense that with any straightforward measure they occupy an exponentially small (and unstable) region of phase space.  If regarded as fluctuations from equilibrium, they are very unlikely.\footnote{As pointed out in \cite{Holman:2005qk}, there is no reason at all to suppose that the universe was ever in equilibrium.  So it isn't quite right to say that low entropy initial conditions are \emph{unlikely}, hence the euphemism `unnatural'.  Nevertheless we would like an explanation for how low entropy initial conditions might have come about.  The situation is similar to explaining the number of generations of particles.  There is no particular reason to say that 3 is unlikely, but it is still a mysterious fact that we would like to derive.}

So the central problem of the arrow of time consists in finding a justification for the so--called \emph{past hypothesis} \cite{Albert:2003kq} --- the assumption that the universe had low entropy at early times.\footnote{A \emph{statistical hypothesis} is also required, to ensure that the initial low entropy condition is not among the pathological few where the entropy gets lower still as time moves forward.  This amounts to postulating a flat measure (or just about any smooth measure) on the microstates in the initial low entropy macrostate, which is already `natural'.  Our aim is to make the past hypothesis just as natural.}

\subsection{Possible solutions}
\ \ \ \ \ 
At the start of the hot big bang era the universe was very smooth, with density perturbations $\delta\rho/\rho\sim 10^{-5}$.  This corresponds to a state of very low entropy, because the generic fate of classical matter in a decelerating universe is to form an inhomogeneous configuration of black holes, i.e. $\delta\rho/\rho\sim 1$.  It is occasionally suggested that inflation adequately explains this fact, since accelerated expansion smooths out inhomogeneities that would otherwise be susceptible to gravitational collapse \cite{Davies:1983nf,Davies:1984qc}.  The reason this argument fails \cite{Page:1985hx,Hawking:1987bi,Hollands:2002yb,Carroll:2005it} is that inflation itself requires extremely special initial conditions to get going.  In the simple case of a single scalar field $\phi$ in a potential $V(\phi)$, the equation of state parameter is given by
\begin{align}
  w = \frac{\frac{1}{2}\left(\dot{\phi}^2+(\nabla\phi)^2\right)-V}{\frac{1}{2}\left(\dot{\phi}^2+(\nabla\phi)^2\right)+V}.
\end{align}
For acceleration, we need $w<-1/3$, which requires the potential to dominate over `kinetic' energy.  Since spatial derivatives contribute to the latter, the initiation of inflation requires an expanse of spacetime to contain an extremely smooth scalar field.  One can see that inflation only makes the problem worse in another way by considering a generic state of the sort likely to have come about shortly after reheating in an inflationary scenario --- most such states do not `un--reheat' when evolved back in time, establishing the specialness of the initial early universe state.  Similar conclusions apply to the horizon problem inflation purports to solve-- getting inflation started requires more tuning than simply giving similar temperatures to the various causally disconnected patches at decoupling.

Despite arguments that no dynamical solution of the problem is possible \cite{Wald:2005cb}, an interesting scenario was proposed recently by Carroll \cite{Carroll:2004pn}. In this approach the question of why the entropy of the universe was so low in the past becomes meaningless because there is no equilibrium state with maximal entropy, and thus any finite entropy is as (un)natural as any other.  Here we propose an alternative explanation, where the low entropy arises from a phase transition in which the strength of gravity increases.  A cosmological time asymmetry arises  from varying spatial asymptotics that leave the local laws of physics time-reversal invariant. 

\section{Entropy and gravity}
\ \ \ \ \ 
Gravitational interactions alter the macroscopic appearance of what we consider low and high entropy states.  Imagine a weakly interacting gas in a box of fixed finite volume.  In the absence of gravity the gas spreads out.  The equilibrium state (i.e. the state with maximal entropy) is homogeneous and smooth, and we would be puzzled to find the system in an extremely lumpy configuration.

Now consider the same system, but take account of gravity and imagine that the size of the box exceeds the Jeans' length of the gas.  The system is unstable to gravitational collapse and the result is that most states evolve to highly inhomogeneous configurations of black holes --- eventually just a single black hole.\footnote{We neglect for now quantum and semi--classical effects like black hole decay since these take place on a much longer timescale.  If the volume of the system is small enough (or if in a cosmological setting the effective Hubble constant is large enough), such effects can also be neglected for the reason that the black holes will not fully evaporate but will come to equilibrium with their Hawking radiation.}  With respect to gravitational interactions, the equilibrium high entropy states are the lumpy ones, and we would be puzzled to find the system in an extremely smooth state.  

The idea we will explore is a transition between these two cases.  If, at early times, gravity were sufficiently weak (compared to its strength now), then the natural configurations would be homogeneous.  Then, as gravity became increasingly strong, at some point the universe would find itself in an apparently unnatural state of low entropy (with respect to dynamics that include the newly strong gravitational force), despite starting out in a natural high entropy state.  

\begin{figure}[h!]
\begin{center}
\includegraphics[height=4in]{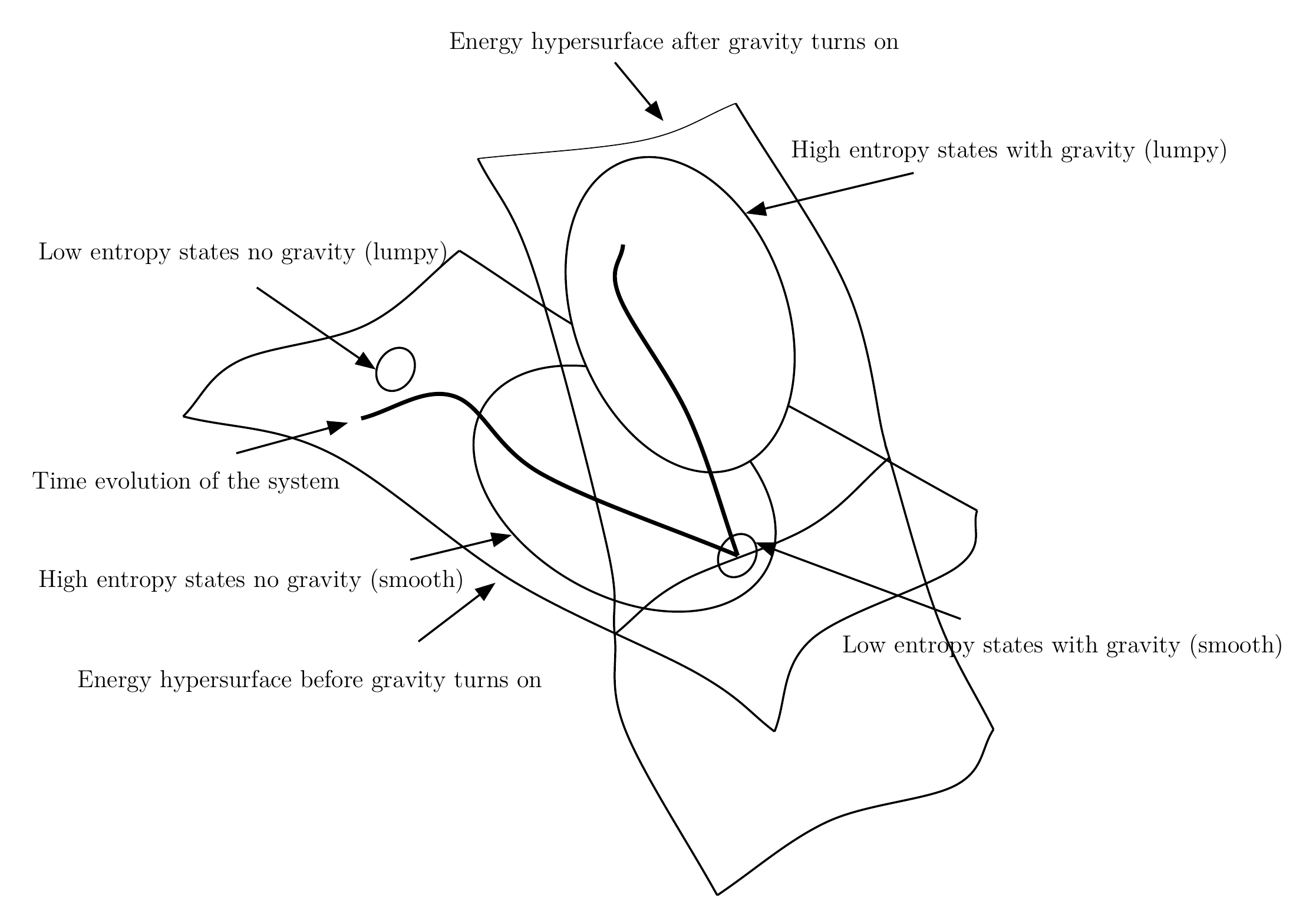}
\caption{Phase space of a gas in a box when gravitational interactions are suddenly turned on.}
\label{phasespace}
\end{center}
\end{figure}

The motion in phase space would look something like that in figure \ref{phasespace}, whose axes represent the positions and velocities of the particles in the box.  Before gravity turns on, the system is confined to a co-dimension one constant $E$ hypersurface, where $E$ is the conserved energy.  This hypersurface is coarse-grained according to the macroscopic appearance of its contents.  Shown is one region containing  clumpy inhomogeneous states, and another with smooth states.  Using a `natural' phase-space measure, the latter occupy significantly more volume than the former, and (again using a natural measure) the vast majority of the lumpy states consist of configurations which subsequently evolve to homogeneous states.

Two things happen when gravity turns on (for ease of discussion, we imagine that gravity turns on instantaneously).  First a negative potential energy is suddenly added to each particle pair.   This changes the numerical value of the energy from $E$ to $E'$, since the positions and velocities (and hence kinetic energies) are left momentarily unchanged.  Second, the foliation of the phase space into constant-energy hypersurfaces is altered, because the potential now allows energy changes with the  momenta of each particle remaining constant.  The result is that the system leaves the original constant $E$ hypersurface and goes onto the constant $E'$ hypersurface it happens to lie on at the time when gravity turns on.   

The $E'$ hypersurface is also coarse-grained into regions according to macroscopic appearance of the states, but now the regions containing smooth states are small in comparison with those containing lumpy states.  The system, since it came from a large entropy region of $E$, necessarily finds itself starting in a tiny smooth region after jumping to $E'$, thus explaining the smoothness of initial conditions in a natural fashion.  In essence, as gravity is turned on, the system is forced from a region of large entropy (with respect to the non-gravitational dynamics) into a region with low entropy (with respect to gravity).  

In this simple example, we imagined that the change in the strength of gravity is accomplished by some external agent.  If we want the strength of gravity to increase dynamically, we need to confront the question of whether the initial state of the dynamical degrees of freedom in the gravitational interaction itself must be fine-tuned to allow the transition to happen.  However, if the time variation in the strength of gravity is truly external, there is no such issue; this amounts to solving the arrow of time problem in a rather straightforward way -- making the laws of physics explicitly time dependent.  

The remainder of this paper develops a simple model that realizes this scenario cosmologically, with one important feature that allows the laws of physics to remain locally time-reversal invariant (as deduced from laboratory experiments).  The explicit time dependence will come not from the lagrangian, but rather from spatial boundary conditions which we impose on the solutions via a choice of vacuum.  The laws of physics are thus locally time reversal invariant, but not so on cosmological scales large enough to be sensitive to the effects of the spatial boundary conditions.

\section{Review of spatial boundary conditions}
\ \ \ \ \
Within the framework of effective field theory, there are three classes of parameters which go into generating predictions:
\begin{itemize}
\item the parameters in the lagrangian,
\item the parameters describing the spatial asymptotics obeyed by the fields, i.e. the choice of vacuum,
\item the parameters describing the initial conditions of the fields, i.e. the choice of state.  
\end{itemize}

The lagrangian specifies the equations of motion obeyed by the degrees of freedom, and for the field theories that describe our universe well at accessible energies, these equations are hyperbolic.  
Hyperbolic equations generally require two kinds of boundary data to guarantee a unique solution.  The first is a choice of spatial asymptotics: unless the spatial sections are compact, the solution should approach some specified, possibly time--dependent, background solution at spatial infinity.  This corresponds to a choice of vacuum.  The second is a choice of field values and momenta over a space-like Cauchy surface, whose spatial values are consistent with the choice of vacuum.  These are the initial conditions, which correspond to a choice of state within the Fock space built over that vacuum\footnote{Of course, in reality there is no hope of knowing the initial conditions precisely --- what one works with is a probability distribution over initial conditions.}.  

A field with a given choice of vacuum, i.e. of spatial asymptotics, cannot dynamically evolve into a field with a different choice;\footnote{Solutions with different boundary conditions are separated from one another by infinite energy barriers.  Quantum mechanical tunneling between vacua is possible via bubble nucleation, but the bubble forms in a background of another vacuum whose asymptotics are unchanged. } indeed, this is part of the fixing of boundary conditions.  The space of all field histories is split into superselection sectors according to the vacuum to which they belong.   About each vacuum, we expand the field in modes.  Expanding the lagrangian and equations of motion around the vacuum will give the dynamics of these modes.\footnote{Note that the space of solutions around a given vacuum include \emph{all} solutions with the given boundary data, not just small fluctuations about the vacuum.  For example, the asymptotically flat sector of GR should include the Schwarzschild black hole.}  

The dynamics of the modes will be generated by some hamiltonian.  For example, for the gravitational modes in GR it is the ADM energy \cite{PhysRev.122.997} ,  
\be E=-{1\over 8\pi}\int _{S_t^\infty}N\left(^{(2)}k-^{(2)}k_0\right)-N^a p_{ab} r^b.\ee
Here the integral is over a 2-sphere at spatial infinity bordering a spatial slice at some fixed time, $^{(2)}k$ is the trace of the extrinsic curvature of the 2-sphere as embedded in the spatial slice, and $^{(2)}k_0$ is the same quantity, though calculated with respect to the background metric.  $N^a$ is the lapse, $N$ the shift, $p_{ab}$ the canonical momentum to the induced metric on the spatial slice, and $r^a$ the outward--pointing normal vector to the 2-sphere tangent to the space-like surface.  

Energy will only be conserved if the background is time translation invariant.  In the case of a non-trivial, time-dependent background, such as an FRW cosmology, $^{(2)}k_0$ will have explicit time dependence, and so the hamiltonian will depend explicitly on time, and will not generate a symmetry.  Even though the equations of motion, and hence the local laws of physics, are time translation (and time reversal) invariant, the spatial boundary conditions break this symmetry.  

We are proposing that spatial boundary conditions represent the `external force' which breaks time reversal invariance in the universe and gives insight into the dynamics that makes smooth early conditions more natural.  Although a standard FRW universe also breaks time reversal invariance in this way, the time-dependence of the hamiltonian is not harnessed in a way as to explain the smooth initial conditions; the mere presence of explicit time-dependence in the metric does not by itself guarantee that the universe will start out with the homogeneous initial conditions that gave rise to our universe 

In what follows, we present a simple model in which there is a background which starts in a weak-gravity situation and evolves to a strong-gravity situation.  If the spatial boundary data are chosen to follow this solution, then we have a situation similar to the gas in the box, and we expect that generic initial states will evolve to give a universe like ours.  

\section{A model for weakening gravity dynamically}
\ \ \ \ \
Our goal is to construct a theory where natural initial conditions lead to a solution in which the local strength of gravity evolves dynamically from being very weak in the early universe to very strong today.  Ideally we would like to have parametric control of the time at which the transition happens, and of the ratio of the gravitational couplings at late and early times.  

We will accomplish this by building a theory of gravity with two maximally symmetric stationary points, one of which has a larger gravitational coupling than the other.  The weak gravity point will be unstable, and initial configurations that start near it will roll down to the second (stable) point, in which the local strength of gravity is much stronger.

\subsection{Brans--Dicke theory}

A simple strategy for arriving at a theory of the kind we need is to promote Newton's constant to a dynamical variable, and give it a suitable potential.  This leads us to consider a scalar--tensor action:
\be
\label{generic}
S = \int d^4x \, \sqrt{-g} \left(f(\phi) R
- {1 \over 2}g(\phi) g^{\mu\nu}\partial_\mu \phi \partial_\nu \phi - U(\phi)\right)+{\cal L}_M\,.
\ee
Given a vacuum solution in which the scalar takes the VEV $\phi_0$, the effective Planck length $\kappa_{\rm eff}$ is the coefficient of the Einstein-Hilbert term
\be {1\over 2\kappa^2_{\text{eff}}}=f(\phi_0).\ee  
Two alternative presentations of \eqref{generic} will be helpful for the subsequent analysis.  The first makes use of the fact that a subset of Brans--Dicke theories can be written as $F(R)$ gravity: 
\be \label{FofRgrav} \int d^4 x \ \sqrt{-g}F(R) + {\cal L}_m(g_{\mu\nu}).\ee
Existence of the two stationary points, and the stability of one of them, as well as the requirement that we get different effective Planck constants in the different vacua, then put constraints on the function $F(R)$, which we'll find can be met by a simple polynomial whose coefficients control the ratio of Planck masses between solutions.  

To see that $F(R)$ actions can be reformulated as Brans--Dicke theories, consider the following action\footnote{See section 5 of \cite{Dyer:2008hb} for more about scalar--tensor/$F(R)$ equivalence and references to the literature, and \cite{Batra:2008cc} for more on engineering theories with given vacua and Planck masses.}
\be \label{scalartens} \int d^4 x \sqrt{-g}\left[F(\phi)+F'(\phi)(R-\phi)\right]+{\cal L}_m(g_{\mu\nu}).\ee
The equation of motion for the scalar is 
\be F''(\phi)\left(R-\phi\right)=0\ee
which implies $R=\phi$ provided $F''\not= 0$.\footnote{Note that the scalar $\phi$ has mass dimension 2.} Plugging this back into the action recovers the original $F(R)$ action, so the two are classically equivalent\footnote{$F(R)$ theory is $\omega=0$ Brans--Dicke theory with a scalar field $F'(\phi)$ and a potential.  It is the potential that allows some $F(R)$ theories to evade solar system tests.  Without it, solar system constraints require $\omega\gtrsim 40,000$.}
. 
A stationary solution in the $F(R)$ theory is a constant curvature solution $R=R_0$ to the equation $F'(R)R-2 F(R)=0$.  The $\phi$ equation of motion sets $\phi=R$, so $\phi$ takes the constant value $\phi_0=R_0$ at the stationary point.  From the Brans--Dicke action \eqref{scalartens} we then read off the effective value of the Planck mass 
\be {1\over 2\kappa^2_{\text{eff}}}=F'(\phi_0)=F'(R_0).\ee  

So far we have presented the theory in two ways: an $F(R)$ action, and an equivalent Brans--Dicke theory.  It will be useful to consider one further action, found by performing a conformal transformation on the Brans--Dicke action to put the Einstein--Hilbert term into canonical form.  As long as $F'(\phi)\not=0$, one can make a conformal transformation to a new metric $\tilde{g}_{\mu\nu}$
\be \tilde{g}_{\mu\nu}=2\kappa^2 F'(\phi)g_{\mu\nu}.\ee
The (at this point arbitrary) constant $\kappa^2$ will have to be positive or negative according to whether $F'(\phi)$ is positive or negative, and separate conformal transformations may have to be made for separate regions in field space.  

The resulting theory reads 
\be \label{EHaction} \int d^4 x \  \sqrt{-\tilde g}\left[ {1\over 2\kappa^2}\tilde R-\half K(\phi)(\partial\phi)^2-V(\phi)\right],\ee
where
\be K(\phi)= {1\over 2\kappa^2} {3F''^2\over F'^2},\ \ \ V(\phi)={\phi F'-F\over (2\kappa^2)^2 F'^2}.\ee
Whether the scalar is normal or ghost-like depends only on the sign of $\kappa^2$, which is the same as the sign of $F'$ in the region under consideration.  

Once the conformal transformation is made the scalar $\phi$ is minimally coupled to gravity, so vacuum solutions and their stability can be straightforwardly read off from the potential $V(\phi)$.  One can check that the points $\phi_0$ at which $V'(\phi_0)=0$ correspond to curvatures $R_0=\phi_0$ satisfying $F'(R_0)R_0-2 F(R_0)=0$ in the original $F(R)$ theory.  The second derivative $V''(\phi)$ tells us about the stability of our vacua.  Evaluated on stationary solutions, we have 
\be V''(\phi_0)=\frac{F''(\phi_0 ) \left(F'(\phi_0 )-\phi_0  F''(\phi_0 )\right)}{(2 \kappa ^2)^2 F'(\phi_0 )^3}. \ee
However, one must keep in mind that in this frame matter is not minimally coupled.  Matter particles do not follow geodesics of $\tilde{g}_{\mu\nu}$.   It is more natural to think of the Jordan frame \eqref{generic} as physical.  

To recap, the equivalent forms of the action are \eqref{FofRgrav}, the $F(R)$ action, \eqref{scalartens}, the scalar-tensor form where matter is minimally coupled, and \eqref{EHaction},  the scalar-tensor form with a canonical Einstein-Hilbert term.  In the first, the simplicity of the $F(R)$ action facilitates the search for the members of this class of theories with the properties we want, the second is useful for studying the behavior of matter at a fixed $\phi$ stationary point, the third makes the dynamics of $\phi$ more transparent.

\subsection*{Constraints on $F(R)$}

With the above in mind, we seek a function  $F(R)$ that satisfies the following criteria:
\begin{enumerate}
  \item $F'(R)R-2 F(R)=0$ should have solutions at $R=0$ and at $R=\lambda$, where $\lambda>0$, and there should be no other solutions between these two. We choose $\lambda>0$ so as to allow the flat FRW ansatz.   We want to start off in the $R=\lambda$ solution and roll down to the $R=0$ solution.  There will be no exponential expansion at the endpoint, and we'll expect solutions to approach a standard power law FRW universe.  
   \item We want $F'(R)>0$ everywhere between the two solutions.  This will ensure that the conformal transformation is valid in this entire region.
  \item We want $F''(R)>0$ everywhere between the two solutions.  This will ensure that the equivalence between \eqref{FofRgrav} and \eqref{generic} is valid everywhere in the region, and also that the effective Planck mass at the high curvature point is larger than at the zero curvature point.  
  \item We want \[V''(0)>0,\] so that the zero curvature solution is stable.  As a consequence of the other conditions, the high curvature solution will automatically be unstable, and since there are no stationary points in between, we can expect that any cosmology starting near the high curvature solution will dynamically evolve towards the low curvature solution.  
    
\end{enumerate}
If we try to satisfy these constraints using a polynomial, we find they cannot be satisfied by a quadratic, or by any polynomial of odd order.  For a quartic, the constraints amount to:
\begin{align}
  F(R) = R + CR^2 + \left( \frac{1}{\lambda^2}-2E\lambda \right)R^3 + ER^4
\end{align}
with either:  $C > 0$ and  $0<E\leq (2\lambda^3)^{-1}$, or $C=0$ and $E= (2\lambda^3)^{-1}$.  We will fix $E$ but allow $C$ to vary:
\begin{align}\label{ourFofR}
  F(R) = R + CR^2 + \frac{R^4}{2\lambda^3}.
\end{align}
The factor by which $G$ then increases as the universe moves from $R=\lambda$ to $R=0$ is:
\be
  \frac{G_{eff}(R=0)}{G_{eff}(R=\lambda)}={\kappa^2_{eff}(R=0)\over\kappa^2_{eff}(R=\lambda))} ={F'(\lambda)\over F'(0)}={3+2C\lambda}.
\ee

\subsection{Our model}
\begin{figure}[h!]
\begin{center}
\includegraphics[height=3in]{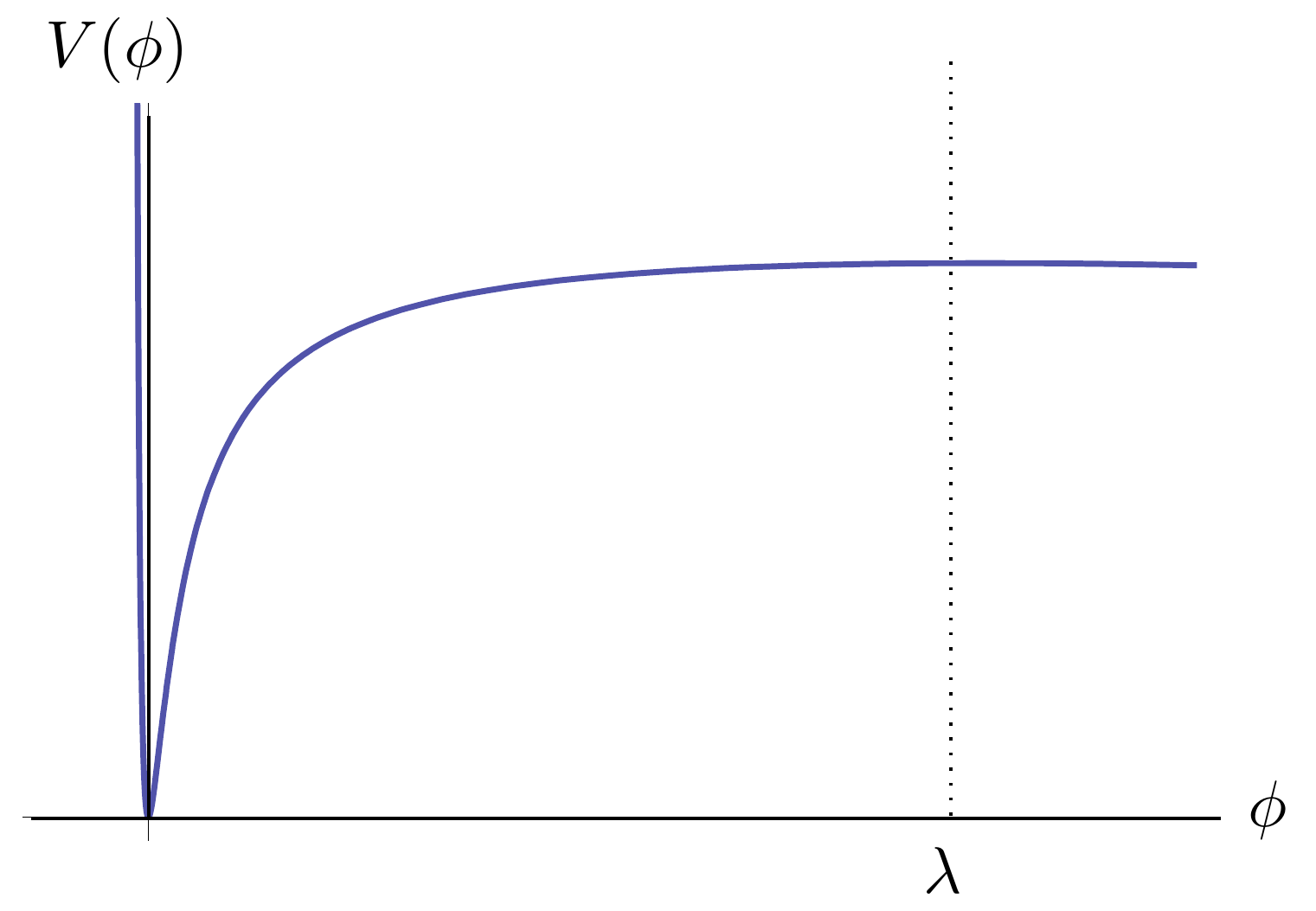}
\caption{Potential in Einstein frame.\label{EinsteinFramePotential}}
\label{plot1}
\end{center}
\end{figure}

With our choice of $F(R)$ \eqref{ourFofR} the action \eqref{FofRgrav} becomes
\be  \label{our_model} \int d^4 x \sqrt{-g}{1\over 2\kappa^2}\left[ R + CR^2 + \frac{R^4}{2\lambda^3}\right]+{\cal L}_m(g_{\mu\nu}).\ee
The equivalent scalar tensor model is 
\be \int d^4 x \sqrt{-g}\left[f(\phi)R-U(\phi)\right]+{\cal L}_m(g_{\mu\nu}),\ee
where 
\bea f(\phi)&=&F'(\phi)={1\over 2\kappa^2}\left(1+2 C \phi+ \frac{2 \phi ^3}{\lambda ^3}\right),\\
U(\phi)&=&\phi F'(\phi)-F(\phi)={1\over 2\kappa^2}\left(C \phi ^2+\frac{3 \phi ^4}{2 \lambda ^3}\right).
\eea
The potential in Einstein frame (see figure \ref{EinsteinFramePotential}), from which we can read off the dynamics, is 
\be V(\phi)={\phi F'-F\over (2\kappa^2)^2 F'^2}=\frac{\lambda ^3 \left(3 \phi ^4+2 C \lambda ^3 \phi ^2\right)}{4 \kappa^2 \left((2 C \phi +1) \lambda ^3+2 \phi ^3\right)^2}.\ee 

\subsection*{Dynamics}

\begin{figure}[h!]
\begin{center}
\includegraphics[height=3in]{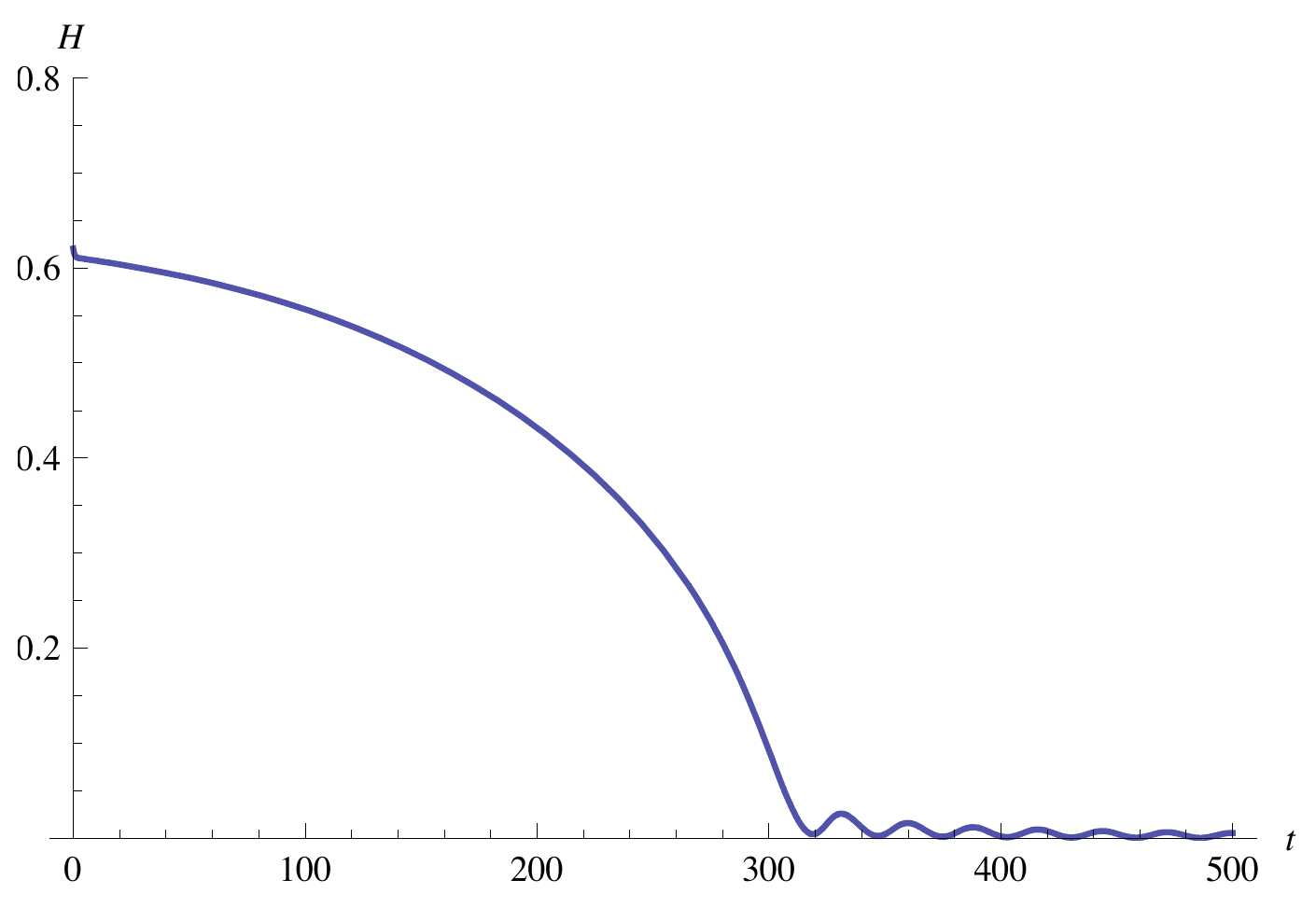}
\caption{Hubble parameter.\label{fig_Hubble}}
\label{plot1}
\end{center}
\end{figure}

\begin{figure}[h!]
\begin{center}
\includegraphics[height=3in]{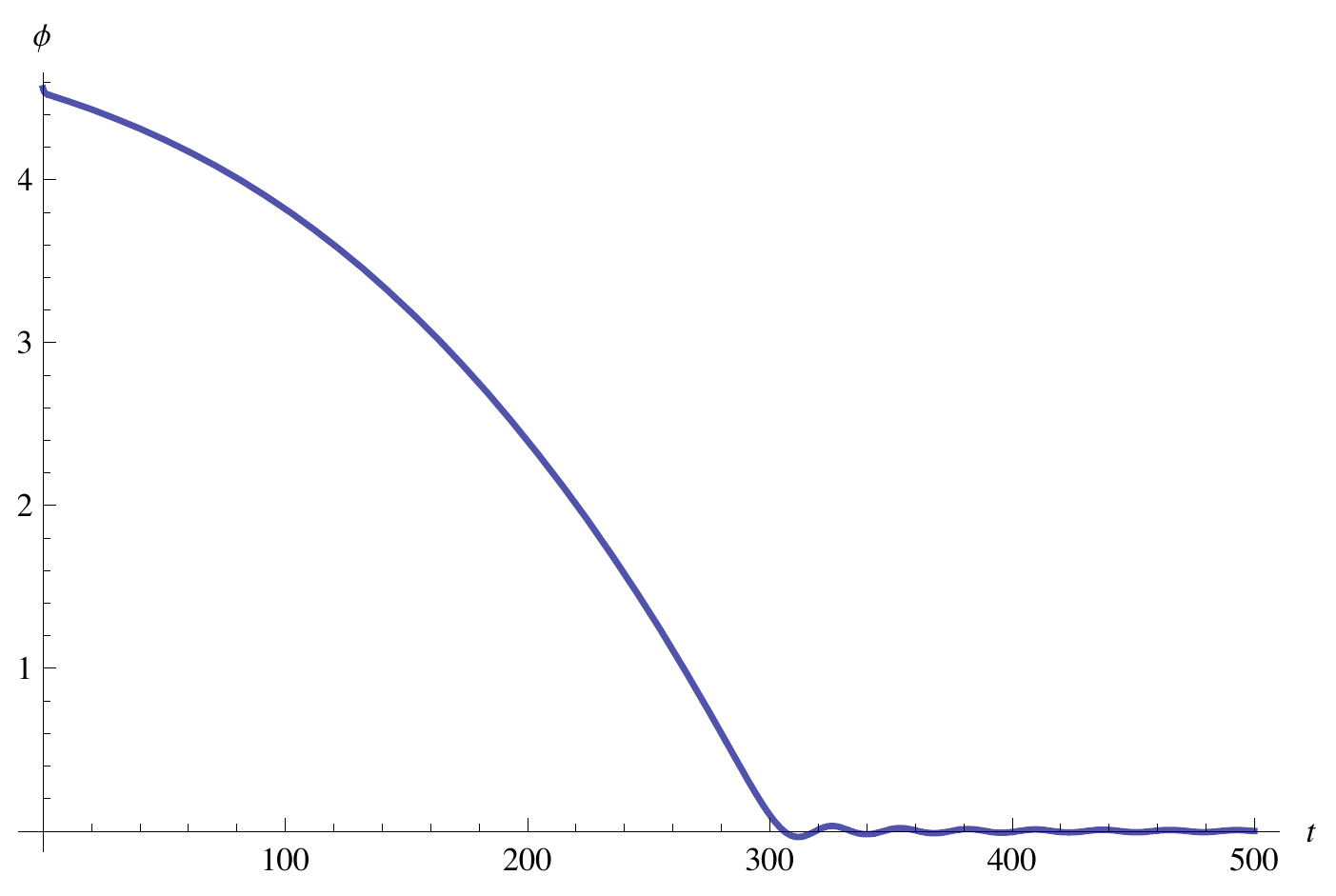}
\caption{Scalar VEV.\label{fig_scalar}}
\label{plot1}
\end{center}
\end{figure}

\begin{figure}[h!]
\begin{center}
\includegraphics[height=3in]{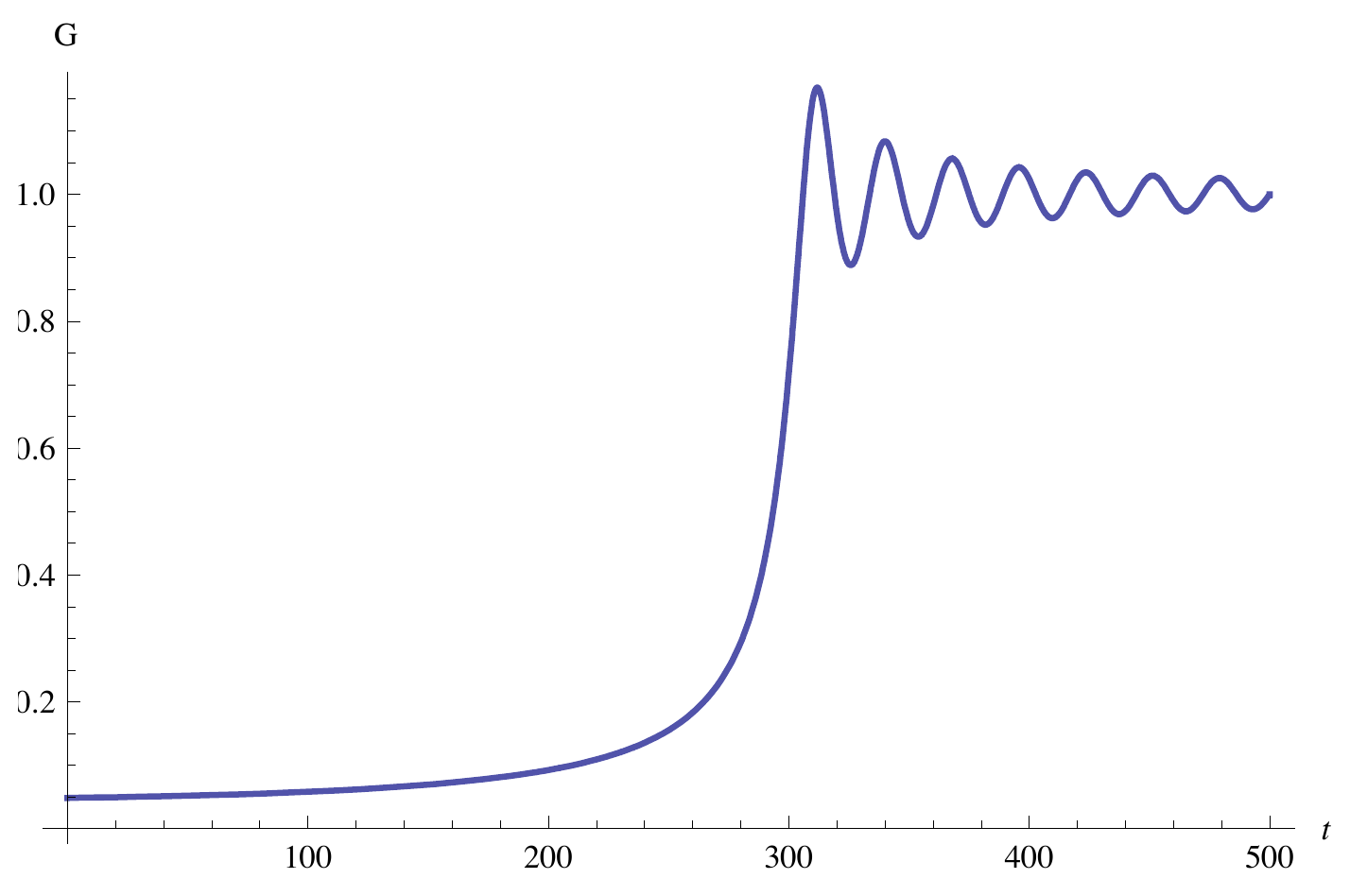}
\caption{Effective $G$.\label{fig_effectiveG}}
\label{plot1}
\end{center}
\end{figure}

Figures \ref{fig_Hubble}--\ref{fig_effectiveG} show the time evolution in the model $\lambda = 5$, $C=2$, with initial conditions  $\phi = 5$, and $\phi$ given an initial velocity $\sim -1$.  The Hubble parameter starts out approximately constant, as would be expected in an almost stable point with $R>0$, and then descends toward zero as the spacetime approaches a power law FRW universe.  Figure \ref{fig_scalar} shows the scalar rolling towards the origin during the transition, and figure \ref{fig_effectiveG} shows the consequent behavior of the effective gravitational coupling (normalized so that $G_{\text{present day}}=1$) which increases by a factor of $3+2C\lambda=23$.  

The matter content has been set to zero, but the picture remains qualitatively unchanged after adding matter.  In fact, adding matter prolongs the time spent near the unstable stationary point.  

We can gain parametric control over when the transition happens by adding a scalar kinetic term to the Jordan frame action, 
\be \int d^4 x \sqrt{-g}\left[f(\phi)R-\half h(\partial\phi)^2-U(\phi)\right]+{\cal L}_m(g_{\mu\nu}),\ee
where $h$ is a constant.  Adding this term does not alter the location, stability, or physical properties of the vacua, but by making $h$ large, we damp the motion of $\phi$, prolonging the time it spends near the unstable stationary point.\footnote{The additional kinetic term also suppresses spatial fluctuations of $\phi$.}  

The numerical simulations above show that large increases in the value of Newton's constant can be brought about in the model \eqref{our_model}.  We therefore suggest that a scenario like this can naturally account for the observed smoothness of the CMB and the low entropy initial conditions of our universe.

There is of course fine tuning in the lagrangian and in the spatial boundary conditions.  Indeed, $F(R)$ lagrangians are very unnatural from a quantum field theory viewpoint, and there is no a priori preferred reason to choose spatial boundary conditions as we have.  We therefore don't claim to have found a completely natural and tuning--free solution to the arrow of time problem, or a complete theory of the early universe, but only the more modest goal of giving a concrete model in which the \emph{initial conditions} can appear highly tuned, but aren't.   

\section{Discussion and possible issues}
\ \ \ \ \
The road to a natural explanation of the low entropy initial conditions of the universe is fraught with difficulties, and one must be particularly sensitive to implicit assumptions tantamount to the conclusion itself, which thereby render the putative solution question--begging.  The subtlety of such considerations makes it difficult to confidently declare a given scenario to be free of fine--tunings.  Several apparently questionable assumptions were made in the model above, some of which we discuss here, mindful of the fact that others may be lurking undetected.

The simulations were carried out with a homogeneous scalar field.  This serves only to identify the background solution setting the spatial asymptotics.  One must imagine a set of initial fluctuations on top of these solutions.   Indeed, if we are allowed to choose homogeneous initial conditions then the very problem we are addressing --- smooth initial condition --- evaporates.  One might worry that the increase in the strength of gravity disappears with more general initial conditions.  However, we argue that the homogeneity of $\phi$ is not necessary in the above scenario (except for computational convenience).  The phenomenon we rely on is just the rolling of the expectation value $\left<\phi\right>$, and unlike inflation, this does not require a smooth field.   Our background solution, which is indeed fine tuned (it starts near the unstable point, whereas it could have started anywhere), only serves to set the spatial asymptotics, and allows us to argue that generic configurations of initial fluctuations will lead to a universe like ours.  We do not address how inflation may tie in with the weakening of gravity, or whether it can be realized in a model like this, since this would obfuscate our main point.

Our scenario does pass one crucial test, avoiding the ``double standard'' \cite{Price:1993hr}.  Often natural--seeming scenarios, such as a ball rolling down a hill and coming to a stop due to friction, become completely unnatural viewed in time reverse.  In our scenario, the time reverse is completely natural and goes as follows:  gravity is strong and the universe starts in a clumpy state.  Then at some point gravity turns off.  The clumps start to dissolve into a smooth state.  The state at all times is natural with respect to the laws operating at those times.  Passing the double standard test requires that the universe spend a sufficient period of time in the weak gravity phase, otherwise there will not be sufficient time (in the time reversed sense) for a generic initial clumpy state to smooth out.  In our model, we accomplish this by tuning the parameter $h$ in the lagrangian.  As we've mentioned, the parameters in the lagrangian are fine--tuned; it is only the initial conditions that are allowed to be generic. 

Again, our model itself is meant only to provide a concrete realization of the underlying idea of weakening gravity, not to be a complete theory of the early universe.  As such, the idea of an early phase characterized by weaker gravity should be kept distinct from the particular features of the model presented.  Indeed, our particular model (making use of spatial boundary conditions as a time dependent driving force) may not be the only or best approach to weaken gravity in the early universe.  Another class of models that may do the job is presented in \cite{Biswas:2005qr}.  The essential, model independent, point is that  a gravitational theory whose strength was sufficiently weak at early times can make smooth initial data the norm, and so potentially offers a different perspective on the the usual puzzle of low entropy initial conditions.

\subsection*{Acknowledgments}
\ \ \ \ \
The authors wish to thank David Albert, Sean Carroll, Ethan Dyer, Dan Kabat, Janna Levin and Ali Masoumi for discussions, and acknowledge support by DOE grant DE-FG02-92ER40699.  BG and MP were supported in part by FQXi grant Arrows of Time in the Quantum Universe.

\end{document}